\documentclass{cjaa}                   

\usepackage{graphicx}                   
\def\ga{\lower.4ex\hbox{$\;\buildrel >\over{\scriptstyle\sim}\;$}}
\def\la{\lower.4ex\hbox{$\;\buildrel <\over{\scriptstyle\sim}\;$}}

\begin{document}

   \title{What is Special about HBRPs
}

   \volnopage{Vol.0 (200x) No.0, 000--000}      
   \setcounter{page}{1}          

   \author{N. Vrane\v{s}evi\'{c},
      \inst{1,2}\mailto{}
D.~B. Melrose  
      \inst{2}
\and    R.~N. Manchester
      \inst{1}}
   \offprints{N. Vrane\v{s}evi\'{c}}                   

   \institute{ATNF, CSIRO, Epping, NSW 1710, Australia\\
             \email{natasa.vranesevic@csiro.au}
        \and
             School of Physics, University of Sydney, NSW 2006, Australia\\
          }

   \date{Received~~2006 month day; accepted~~2006~~month day}

   \abstract{ The Parkes Multibeam Survey led to the identification of
a number of long-period radio pulsars with magnetic field well above
the `quantum critical field' of $\sim 4.4 \times 10^{13}$ G
(HBRPs). Traditional pulsar emission theories postulate that radio
emission is suppressed above this critical field. The aim of this
project is to understand emission properties of HBRPs.
 }

   \authorrunning{N. Vrane\v{s}evi\'{c}, D.~B. Melrose \& R.~N. Manchester }            
   \titlerunning{What is Special about HBRPs}  

 \maketitle

%
%

\section{Introduction}           
\label{sect:intro}

About 1500 neutron stars have so far been detected in the Galaxy as
radio pulsars, of which 125 are millisecond pulsars (from ATNF Pulsar
Catalogue at http://www.atnf.csiro.au/research/pulsar/psrcat, see
Manchester et al.~\cite{manc05}). The Parkes Multibeam Survey
(hereafter PMS) doubled the number of known pulsars (Manchester et
al.~\cite{manc01}; Morris et al.~\cite{morr02}; Kramer et
al.~\cite{kram03}; Hobbs et al.~\cite{hobb04}; Faulkner et
al.~\cite{faul04}). Prior to the PMS the highest measured neutron star
surface magnetic field was $2.1\times 10^{13}\rm\,G$ for the pulsar
B0154+61. The PMS led to the identification of a number of `HBRPs',
radio pulsars with long periods and with surface magnetic fields near
$10^{14}\rm\,G$ (McLaughlin et al.~\cite{mcla04}). Even more exotic
neutron stars, with the surface fields clustered around
$10^{14-15}\rm\,G$, are the Anomalous X-ray Pulsars (AXPs) (Kaspi \&
Gavriil~\cite{kasp04}) and Soft Gamma-rays Repeater (SGRs)
(Kouveliotou~\cite{kouv03}), now believed to be 
magnetars (Duncan \& Thompson~\cite{dunc92}). The fields of these
sources, deduced from their dipole spin down rates ($B_{\rm
surf}=3.2\times 10^{19} (P {\dot P})^{1/2}$), are well above the
quantum critical magnetic field of $B_{\rm c}\sim
4.4\times10^{13}\rm\,G$. Existence of these long period radio pulsars
with surface dipole magnetic field strengths higher than critical,
demonstrates that radio emission can be produced in neutron stars with
$B_{\rm surf} > B_{\rm c}$, despite the prediction of the radio-quiet
boundary below which radio emission should cease (Baring \&
Harding~\cite{bari98}).

The HBRPs may be young objects that form a transition
between normal radio pulsars and AXPs (McLaughlin et
al.~\cite{mcla03}). This suggests that there should be many more HBRPs
than those currently known. It may be that the SGRs, AXPs and HBRPs form a
continuum of magnetic activity, or they might be different
phases/states of a more uniform class of object. In order to investigate
these ideas we initiated a project with the main goal of
understanding the emission properties of HBRPs. We chose a sample of
17 HBRPs, together with 17 low magnetic-field radio pulsars selected
to have similar spin period distributions. We observed these 34 pulsars
at three different frequencies in order to obtain their polarimetric
characteristics as well as to find their flux densities and
spectral indexes. High time resolution data on their individual pulses
together with multi-frequency polarimetric data will provide us with a
wealth of information about their emission process. We will compare
observed characteristics of these two samples of pulsars using the
Parkes radio telescope.

We organize this paper as follows. In section~2 we discuss super
strong magnetic fields briefly, and then summarize the classes of
pulsars. In Section~3 we present some preliminary results from our
multi-frequency observations. In Section~4 we discuss the importance
of understanding the emission properties from HBRPs and how they
differ from normal pulsars. Our conclusion are given in the Section~5.

\section{Neutron star populations}

Neutron stars are strongly magnetized objects, with surface magnetic
fields ranging from $10^{6}\rm\,G$ to $10^{15}\rm\,G$. The slowing
down of pulsar rotation implies magnetic fields of order of
$10^{11}\rm\,G$ to $10^{15}\rm\,G$. The rapid rotation of such fields
is important in generating relativistic particles and radio
emission. Plasma accreting onto neutron stars in X-ray binary systems
is channeled to the magnetic poles by fields ranging from
$10^{8}\rm\,G$ to $10^{13}\rm\,G$.

\subsection{Radio emission in super strong magnetic fields}

The condition that the cyclotron energy equal the rest energy of the
electron corresponds to $B$ equal to the quantum critical field
strength of $B_{\rm c} ={m_{e}^{2} c^{3}}/{\hbar e}=4.4\times
10^{13}\rm\,G$, and fields of this order or larger are said to be
super strong. Most pulsars have $B_{\rm surf}\la 0.1\,B_{\rm c}$, and 
HBRPs have $B_{\rm surf}\ga B_{\rm c}$. Some other negligible or forbidden physical
processes become important in pulsar magnetic fields, including 
birefringence of the vacuum,
splitting of one photon into two, the decay of a single photon into
an electron-positron pair, and the rapid radiative loss of
perpendicular energy by all electrons, so that their motion is
one-dimensional along the magnetic field lines. 

The radio emission from pulsars is attributed to an electron-positron
pair plasma created by one-photon decay into pairs. This is actually a
four-stage process: a) a parallel electric field accelerates primary
particles to extremely high energies, (b) these primaries emit gamma
rays, initially directed nearly along the field lines, (c) as the
photons propagate outward, the curvature of the magnetic field causes
the angle between the photon and field line to increase, (d) when this
angle is large enough, the photon decays into a pair.  Any process
that stops any of the steps (a)--(d) operating effectively can prevent
copious pair production, so that the pulsar is `dead' as a radio
emitter. The death-line is usually attributed to the parallel electric
field becoming too weak to provide effective primary particles. For
$B_{\rm surf}>B_{\rm c}$ photon splitting can prevent (c) and (d) from
operating effectively: a photon splits into two lower energy photons
before it reaches the threshold for pair creation. Photon splitting is
known to be possible only for photons of one polarization mode of the
birefringent vacuum: photons with the other polarization mode are
forbidden to split by selection rules attributed to Adler
(\cite{adle71}). Baring and Harding (\cite{bari01}) speculated that
this selection rule might not apply for $B_{\rm surf}> B_{\rm c}$, and
that photons of both polarization might split, so that there is no
pair creation for sufficiently high $B$, implying that neutron stars
with $B_{\rm surf}\ga 3\times 10^{13}\rm\,G$ should not be radio
emitters. This prediction appears to be violated by the radio
detection from two AXPs (Malofeev et al.~\cite{malo05}). It also
suggest that HBRPs should not exist.

More recently, Weise and Melrose (\cite{weis06}) showed that Adler's
selection rules continue to apply for $B_{\rm surf}> B_{\rm c}$. Hence
photons in one mode do not experience photon splitting, but can decay
into pairs. We conclude that the existence of HBRPs is consistent with
theory.

\subsection{Neutron stars classes}

Based on observations, neutron stars may be classified as: \\
{\bf Radio active pulsars} are traditional rotational-powered objects,
divided into two main populations: {\it Millisecond
pulsars}, with spin periods centered around $\sim 3\rm\,ms$ and
magnetic fields around $10^{8}\rm\,G$, with different evolutionary
history from normal pulsars; and {\it Normal pulsars} with spin periods
centered around $0.5\rm\,s$ and magnetic fields around
$\sim10^{12}\rm\,G$, thought to be born with short spin periods, to
spin down on a time-scale of $10^{5-6}\rm\,yrs$, and to cease radio
emission after around $10^{7}\rm\,yrs$); with distinguishable third class of
 {\it HBRPs}, which have
similar spin parameters (long periods and high slowdown rates) as
AXPs, but without detectable high-energy emission. \\
{\bf Radio-quiet isolated neutron stars:} emit thermal-like X-rays,
usually subdivided in four classes: {\it Anomalous X-ray pulsars}, a
small class of solitary pulsars with spin periods in the $6-12\rm\,s$
range, very soft X-ray spectra, and strong magnetic fields of
$10^{14-15}\rm\,G$; {\it Soft Gamma-ray Repeaters}. emitting sporadic,
intense flares of low-energy (soft) gamma rays, with periods and
magnetic field in the similar range as AXPs; {\it Dim Thermal neutron
stars}, which are not associated with supernovae remnants; and {\it
Compact Central sources}, not identified with active radio pulsars or
AXPs/SGRs, and with un-pulsed soft (thermal) emission. AXPs together
with SGRs are believed to be {\it magnetars}, a class of neutron stars
distinct from radio pulsars, in which magnetic energy, rather than the
rotational energy, plays the dominant role in powering the X-ray and
gamma-ray emissions. \\
{\bf Accretion-powered neutron stars} are detected in X-ray binary
systems (more than 200 known) as X-ray pulsars or X-ray bursts powered
by thermonuclear flashes.

\begin{figure}
\centering
\includegraphics[width=105mm,height=120mm]{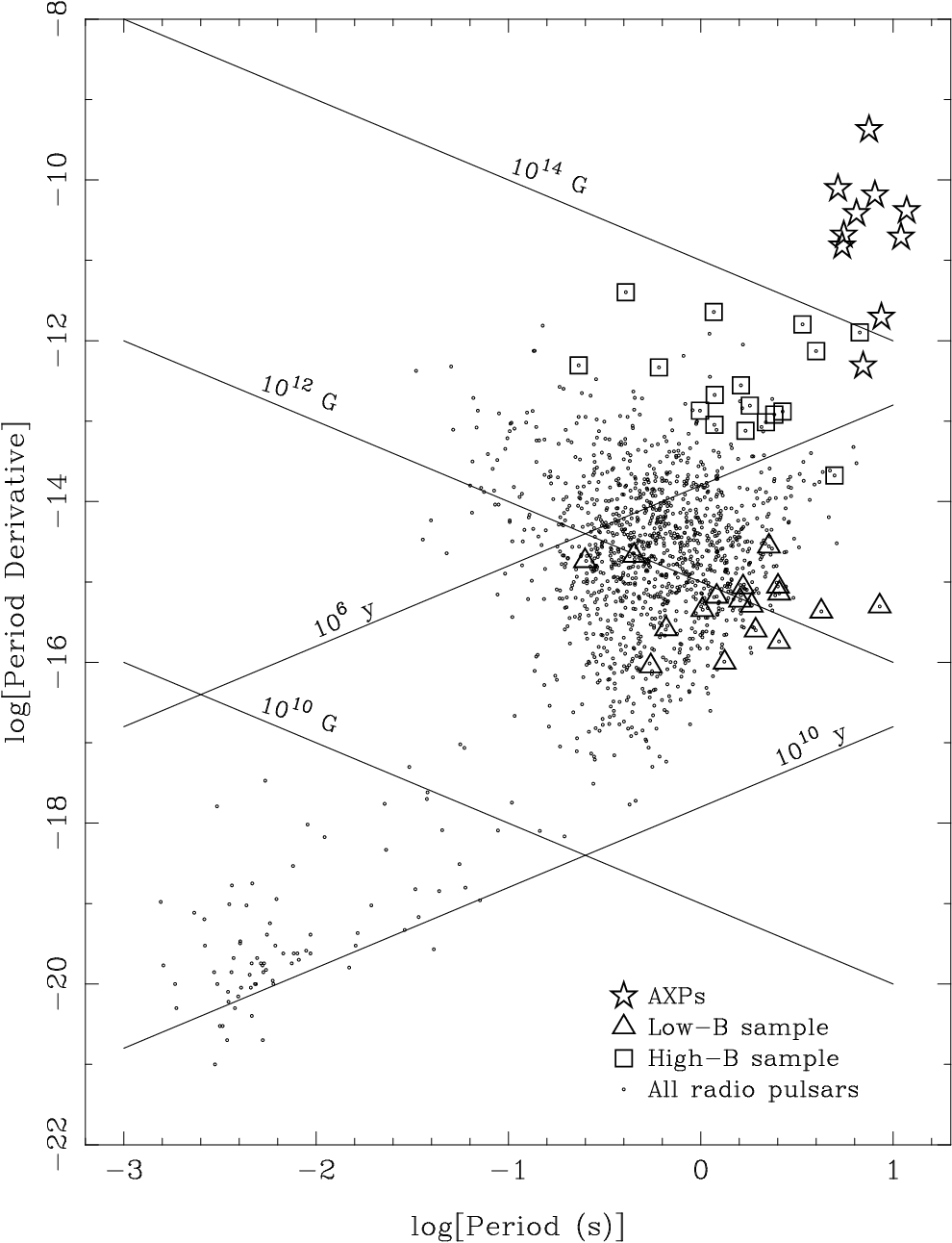}
\vspace{-5mm}
\caption{A $P$ --$\dot P$ diagram showing location of pulsars
populations, as well as the location of pulsars from our samples. Dots
correspond to all radio pulsars taken from the ATNF pulsar
catalogue. Squares and triangles represent pulsars from our high-B and
low-B samples, respectively. AXPs are marked with open stars. Lines of
constant characteristic age and surface magnetic dipole field strength
are shown.}
\label{Fig1:ppdot}
\end{figure}

\section{Some Preliminary Results}

\subsection{Observed samples}

The location of our sample pulsars on the $P - \dot P$ diagram is
shown in Figure 1. Seventeen HBRPs ($B_{\rm surf}> 10^{13}$ G) are
marked with squares. The low-B pulsars sample ($B_{\rm surf}\leq
10^{12}$ G), which has a similar period distribution to the high-B
sample, appear as 17 triangles.  Observations were carried out using
the Parkes 64-m radio telescope, in two sessions, 2005 May 23 -- 25,
and 2005 December 02 -- 05. The 20-cm observations were made using the
central beam of the Parkes multibeam receiver with central
frequencies of 1433 MHz and 1369 MHz. The 10-cm and 50-cm observations
were made using the the dual frequency $10/50\rm\,cm$ receiver with a
central frequencies at 3100 MHz and 685 MHz.  The most exotic pulsars
discovered by PMS are J1718-3718, J1734-3333, J1814-1444, and
J1847-0130, with $B_{\rm surf}\ga B_{\rm c}$, specifically, $7.4\times
10^{13}\rm\,G$, $5.2\times 10^{13}\rm\,G$, $5.5\times 10^{13}\rm\,G$
and $9.4\times 10^{13}\rm\,G$, respectively.

\subsection{PSR J1734-3333}

We show preliminary results for pulsar J1734-3333 at two different
frequencies. This pulsar is young, $\sim 8000\rm\,yrs$ old, with
$P=1.17\rm\,s$ and $B_{\rm surf}>B_{\rm c}$. Figure~2 shows the polarization
profiles at $20\rm\,cm$ and $10\rm\,cm$. The 10cm profile shows two
main components; the dominant trailing component has strong circular
polarization. Linear polarization is significant for both components
with a position-angle swing of opposite sign through the two
components. The 20cm profile clearly shows the effects of ray
scattering by irregularities in the ISM, broadening an 
intrinsically sharp pulse. More distant pulsars with higher DMs are
more likely to be strongly scattered; PSR J1734-3333 has $\rm
DM=578\rm\,cm^{-3}\,pc$ and distance of $d=7.40\rm\,kpc$.

The only publicly available data for this pulsar are filterbank data
at 1374 MHz from the discovery paper (Morris et
al.~\cite{morr02}). More than 80\% of all our observed 34 pulsars have
published data only from their discovery at 20 cm. Detailed results on
multi-frequency observations of HBRPs will be published in the near
future.

\begin{figure}[h]
  \begin{minipage}[t]{0.5\linewidth}
  \centering
  \includegraphics[width=70mm,height=100mm]{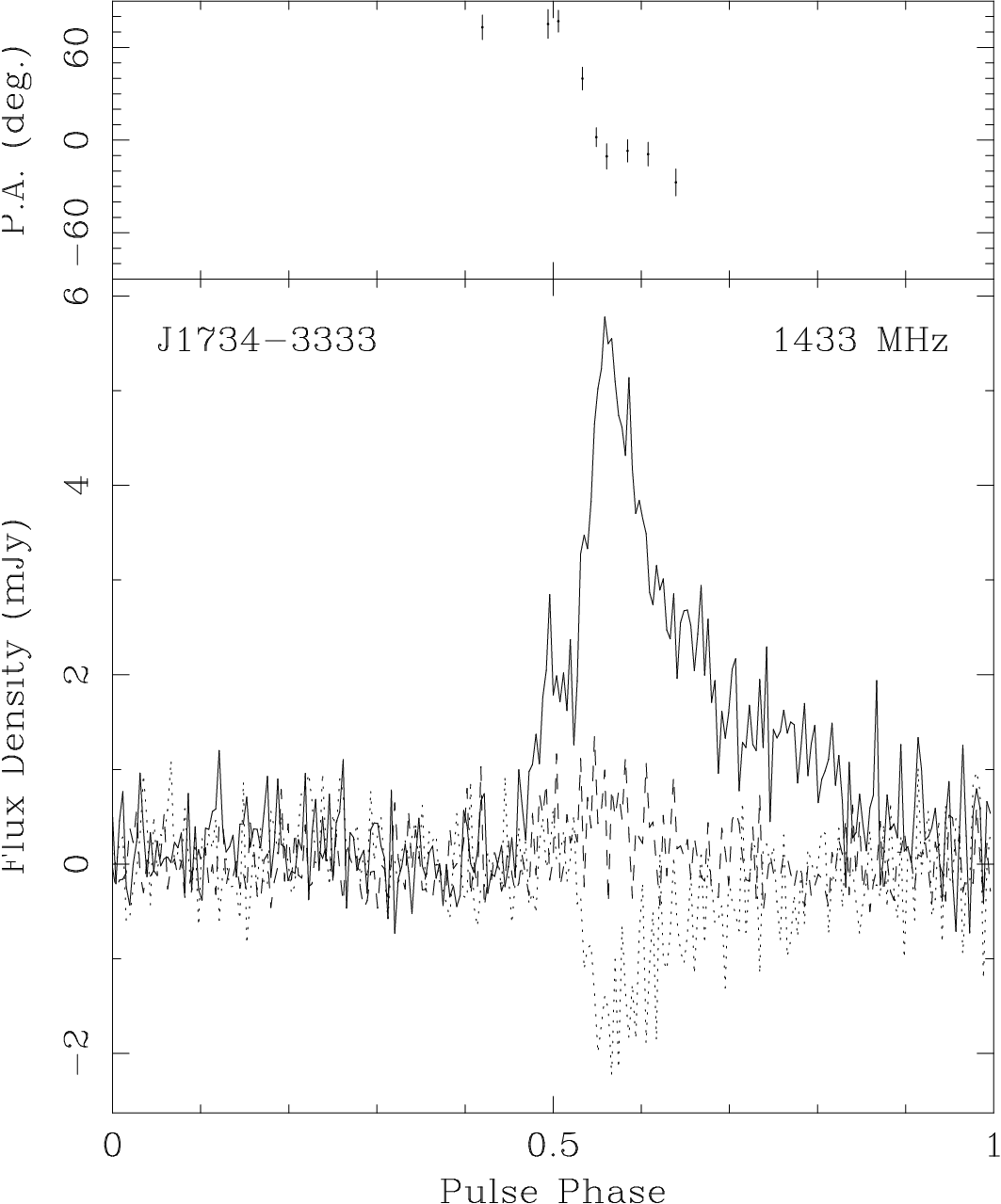}
  \vspace{-5mm}
  \end{minipage}
  \begin{minipage}[t]{0.5\textwidth}
  \centering
  \includegraphics[width=70mm,height=100mm]{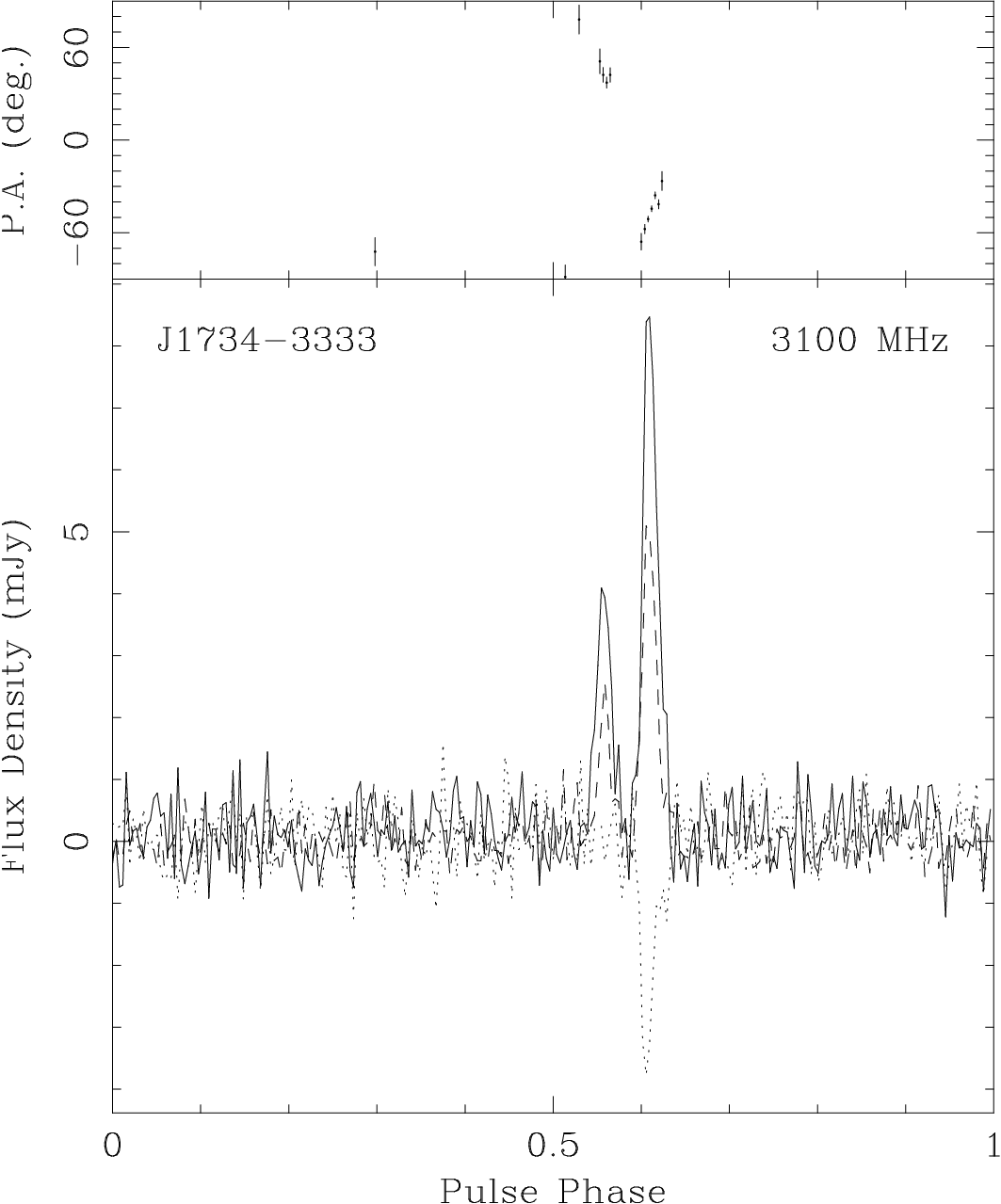}
  \vspace{-5mm}
  \end{minipage}
 \vspace{-5mm}
  \caption{Polarization profiles for PSR J1734-3333 at
  $20\rm\,cm$ and $10\rm\,cm$, showing $360^{\circ}$ of rotational
  phase, with total intensity as a solid line, linearly
  polarized intensity as a dashed line and circularly polarized
  intensity as a dotted line. The upper panel shows the position angle
  of the linear polarization.}
  \label{Fig2:polarprof}
\end{figure}

\section{Discussion}
\label{sect:discussion}

The motivation for this project arose from the desire to
understand recent results of pulsar population analysis (Vranesevic et
al.~\cite{vran04}) which show that pulsars with high magnetic fields
contribute almost half to the total pulsar birthrate, despite the fact
that such high field pulsars are restricted to only few per cent of
the total pulsar population. Furthermore, in the same paper, it was
shown that up to $40\,\%$ of all pulsars are born with periods in the
range $100-500\rm\,ms$, which is in contradiction to the canonical
view that all pulsars are born as fast rotators ($P_{0}\leq
100\rm\,ms$).  Kaspi and McLaughlin (\cite{kasp05}) pointed out the
overlap area (below $10^{14}\rm\,G$), where a couple of magnetars have
fields and periods that are comparable to those of HBRPs. This overlap
suggests that there could exist lower field magnetars (with $B_{\rm
surf}<10^{14}\rm\,G$) that will evolve into X-ray silent HBRPs
(Ferrario and Wickramasinghe~\cite{ferr06}). The same authors have
indicated possibility of discovery of `hybrid' young objects having
these high magnetic fields, exhibiting both magnetar and radio pulsar
characteristics.

\section{Conclusions}
\label{sect:conclusion}

Although HBRPs and magnetars have similar spin parameters, their
emission properties are different. It has been suggested that
pulsars-like objects could evolve from normal radio pulsars to
magnetars  (Lyne~\cite{lyne04}, Lin \&
Zhang~\cite{linz04}). Ferrario and Wickramasinghe
(\cite{ferr03},~\cite{ferr06}) argued that the initial neutron stars
spin periods may depend critically on their magnetic fields, in
particular, there is a tendency for high field systems to be born as
slow rotators. These suggestions may reveal a missing link between
radio pulsars and magnetars.

We hope to provide solid constraints on HBRP's radio emission
characteristics by comparing HBRPs properties with the properties of
normal pulsar population using their multi-frequency and high time
resolution data.
     
\begin{acknowledgements}
This work was supported by the Astronomical Society of Australia (ASA) through Student Travel Funds.
\end{acknowledgements}

\label{lastpage}

\end{document}